\documentclass{ar2e-single}
\usepackage{hyperref} 
\usepackage{graphicx,color}
\usepackage{amsmath}
\usepackage{booktabs}

\newcommand{\bra}[1]{\mathinner{\langle{#1}|}}
\newcommand{\ket}[1]{\mathinner{|{#1}\rangle}}

\newcommand{\braket}[2]{\langle #1|#2\rangle}
\providecommand{\abs}[1]{\lvert#1\rvert}

\begin{document}


\title{Simulating Chemistry using\\ Quantum Computers}

\markboth{Kassal,$^*$ Whitfield,$^*$ Perdomo-Ortiz, Yung, Aspuru-Guzik}{Simulating Chemistry using Quantum Computers}

\author{Ivan Kassal,$^*$ James D. Whitfield,$^*$ Alejandro Perdomo-Ortiz, Man-Hong Yung, Al\'{a}n Aspuru-Guzik\\
\affiliation{Department of Chemistry and Chemical Biology\\ Harvard University, Cambridge MA, United States\\~\\
$^*$ These authors contributed equally to this work.\vspace{-1cm}}}

\begin{abstract}

The difficulty of simulating quantum systems, well-known to quantum chemists, prompted the idea of quantum computation. One can avoid the steep scaling associated with the exact simulation of increasingly large quantum systems on conventional computers, by mapping the quantum system to another, more controllable one. In this review, we discuss to what extent the ideas in quantum computation, now a well-established field, have been applied to chemical problems. We describe algorithms that achieve significant advantages for the electronic-structure problem, the simulation of chemical dynamics, protein folding, and other tasks. Although theory is still ahead of experiment, we outline recent advances that have led to the first chemical calculations on small quantum information processors.

\end{abstract}

\maketitle

\section{INTRODUCTION}

One of the greatest challenges in quantum chemistry is to fully understand the complicated electronic structure of atoms and molecules. Over the last century, enormous progress has been made in describing the general behavior of relatively simple systems. In particular, combined with physical insights, elegant computational approaches, ranging from wavefunction methods to quantum Monte Carlo and density functional theory, have been developed. The challenge is that the Hilbert spaces of quantum systems grow exponentially with system size.  Therefore, as these methods are extended to higher accuracy or to larger systems, the computational requirements become unreachable with current computers. This problem is not merely a consequence of technological limitations, but stems from the inherent difficulty of simulating quantum systems with computers based on classical mechanics. It is therefore important to know if the computational bottlenecks of classical computers can be solved by a computing model based on quantum mechanics---quantum computation---whose development has revolutionized our understanding of the connections between computer science and physics.

The idea of mapping the dynamics of a quantum system of interest onto the dynamics of a controllable quantum system was proposed in 1982 by Feynman \cite{Feynman1982} and developed in 1996 by Lloyd \cite{Lloyd1996}. Such a quantum computer would be able to obtain information inaccessible with classical computers. Consequently, quantum simulation promises to be a powerful new tool for quantum chemistry. In this article, we review the recent applications of quantum simulation to chemical problems that have proven difficult on conventional computers. After introducing basic concepts in quantum computation, we describe quantum algorithms for the exact, non-adiabatic simulation of chemical dynamics as well as for the full-configuration-interaction treatment of electronic structure. We also discuss solving chemical optimization problems, such as lattice folding, using adiabatic quantum computation. Finally, we describe recent experimental implementations of these algorithms, including the first quantum simulations of chemical systems.

\section{QUANTUM COMPUTATION}\label{sec:qc}

\subsection{Differences between quantum and classical computation}

There are fundamental differences between quantum and classical computers. Unlike the classical bit, which is always either a `0' or a `1', the basic unit of quantum information is the qubit (Fig.~\ref{fig:gates}), which can be in a superposition of $\ket{0}$ and $\ket{1}$: $\alpha\ket{0}+\beta\ket{1}$. States of $n$ qubits are elements of an exponentially large, $2^n$-dimensional, Hilbert space, spanned by a basis  of the form $\ket{x_1}\cdots\ket{x_n}\equiv\ket{x_1\ldots x_n}$, where each $\ket{x_i}$ is $\ket{0}$ or $\ket{1}$. This enables entanglement, a feature necessary for the advantage of quantum computers. As an example of entanglement, the two-qubit state $\ket{\Phi^+}=(\ket{00}+\ket{11})/\sqrt{2}$, one of the Bell states, can't be written as a product state $\ket{\phi_1}\ket{\phi_2}$.

The linearity of quantum theory implies that a quantum computer can execute classical computations in superposition. For example, if the input state contains all possible input values $\ket{x}$ of a function $f(x)$, the function can be computed using a unitary operation $U_f$ as
\begin{equation}
\sum_x a_x \ket{x}\ket{0} \stackrel{U_f}{\longrightarrow} \sum_x a_x \ket{x}\ket{f(x)}.\label{eq:superposition}
\end{equation}
With a single call to $U_f$, the quantum computer produces a state that contains information about all the possible outputs of $f(x)$.

Nevertheless, quantum computation has several limitations. For example, the no-cloning theorem \cite{Nielsen2000, Kaye2007} states that an unknown quantum state cannot be copied perfectly. More importantly, the information of a general quantum state cannot be read out with a single projective measurement, because that would collapse a superposition into one of its components. Therefore, while the state in Eq.~\ref{eq:superposition} contains information about all possible outputs, that information is not immediately accessible. Instead, a quantum algorithm has to be designed in a way that makes it easy to measure a global property of $f$, without making it necessary to compute all the individual $f(x)$. Algorithms of this kind are discussed in the following sections.

\subsection{Approaches to quantum computing}

There are several models, or ways of formulating, quantum computation. Most work in quantum simulation has been done in the circuit and adiabatic models. While the two are known to be computationally equivalent---any computation that can be performed in one model can performed in the other in a comparable amount of time \cite{aharonov2007, kempe2004, Mizel2007}---different problems are solved more naturally in different models. We discuss the two models in turn, but note that other models hold promise for the development of future simulation algorithms, including topological quantum computing \cite{kitaev_fault-tolerant_2003,nayak_non-abelian_2008}, one-way quantum computing \cite{raussendorf_one-way_2001,raussendorf_measurement-based_2003}, and quantum walks \cite{kempe_quantum_2003}.

\subsubsection{The circuit model}

The cornerstone of quantum computation is the generalization of the classical circuit model, composed of classical bits and logical gates, to a quantum circuit model \cite{Deutsch1989,Barenco1995,Nielsen2000}. A quantum circuit is a multi-qubit unitary transformation $U$ which maps a set of initial states to some final states. Usually, a unitary gate is decomposed into elementary gates which involve a few (one or two) qubits each.

\begin{figure}[htp]
\begin{center}
\includegraphics[width=10 cm, bb= 150 180 700 300]{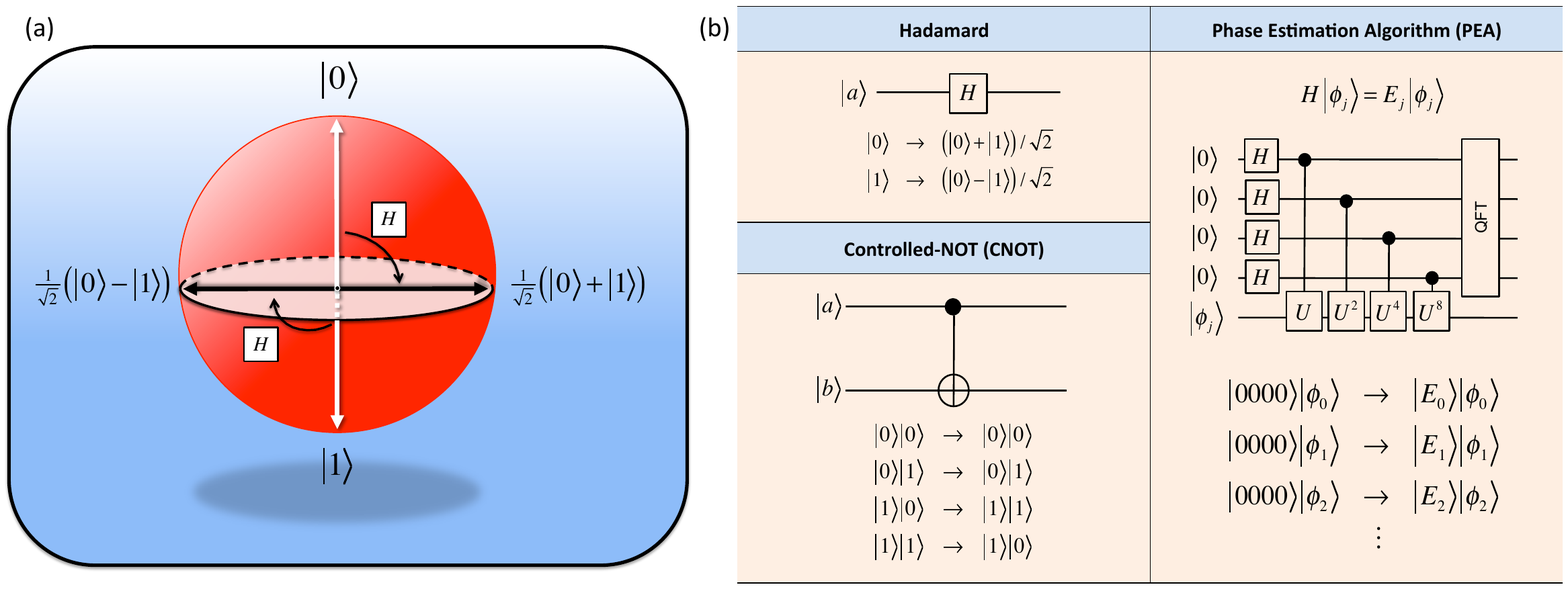}
\end{center}
\caption{\textbf{Qubit, elementary gates and phase estimation algorithm:} (a) The quantum state of a qubit can be represented in a Bloch sphere. (b) The action of the Hadamard gate $ H$ on a qubit is shown in (a). The CNOT gate, together with single qubit gates, form a universal gate set. The quantum circuit for the phase estimation algorithm PEA is shown on the right. Here $U\left| {\phi _k } \right\rangle  = e^{2\pi iE_k } \left| {\phi _k } \right\rangle $ and QFT is the quantum Fourier transform. The eigenvalues  in this case have 4-digit accuracy.}
\label{fig:gates}
\end{figure}

In classical computing, the {\sc nand} gate is universal \cite{null2003}, meaning that any logical circuit can be constructed using {\sc nand} gates only. Similarly, in quantum computing, there are sets of unitary operations that form universal gate sets. A quantum computer that can implement such a set is called universal, and can perform any unitary transformation $U$ to an arbitrary accuracy. It turns out that the set containing all single-qubit gates in addition to any two-qubit entangling gate, such as {\sc cnot}, is universal \cite{Kaye2007} (Fig.~\ref{fig:gates}). An entangling gate can be realized by any physical interaction that can generate entanglement between qubits. Examples of experimental implementations of quantum gates have been reviewed \cite{Ladd2010}, and we will cover some of the experiments relevant to quantum simulation in Sec.~\ref{sec:expt}.

Beside the elementary gates, an important quantum transformation is the quantum Fourier transform (QFT). It transforms any quantum state $\ket{\varphi}  = \sum\nolimits_x {\varphi \left( x \right)} \ket{x}$  into its Fourier representation,
\begin{equation}
U_{QFT} \left| \varphi  \right\rangle  = \sum\limits_{k=0}^{N-1} {\tilde \varphi \left( k \right)} \left| k \right\rangle,
\end{equation}
where $\tilde \varphi(k) = (1/\sqrt N )\sum\nolimits_{x=0}^{N-1} \varphi(x) e^{2\pi ikx/N}$ are the discrete Fourier coefficients of $\varphi(x)$.  The QFT can be efficiently implemented using a quantum circuit~\cite{Nielsen2000}: for $n$ qubits, the number of elementary gates required is $O(n^2)$. For comparison, the classical fast Fourier transform requires $O(n2^n)$ gates. We take advantage of the QFT in Sec.~\ref{ssec:First} for the simulation of quantum dynamics, and in Sec.~\ref{ssec:Measuring} for the measurement of observables.

\subsubsection{Adiabatic quantum computation}\label{sec:adiabatic_intro}

An alternative to the gate model is the adiabatic model of quantum computation \cite{farhi2000}. In this model, the quantum computer remains in its ground state throughout the computation. The Hamiltonian $H(t)$ of the computer is changed slowly from a simple initial Hamiltonian $H_i$ to a final Hamiltonian $H_f$ whose ground state encodes the solution to the computational problem. The adiabatic theorem states that if the variation of the Hamiltonian is sufficiently slow, the easy-to-prepare ground state of $H_i$ will be transformed continuously into the ground state of $H_f$.  It is desirable to complete the evolution as quickly as possible; the maximum rate of change is mostly determined by the energy gap between the ground and first excited states during the evolution~\cite{Messiah,amin_inconsistency_2008,tong_sufficiency_2007,tong_quantitative_2010}. The applications of adiabatic quantum computation to simulation include preparing quantum states of interest and solving optimization problems such as protein folding~\cite{Perdomo2008}. We discuss the details in Secs.~\ref{ssec:ground} and \ref{sec:Adiabatic}, respectively.

\subsection{Quantum complexity theory}\label{ssec:complexity}

To understand the computational advantages of quantum algorithms for chemical simulation, we discuss some aspects of computational complexity theory, which defines quantum speed-up unambiguously. A proper measure of the complexity of an algorithm is how many operations (or how much time) it takes to solve problems of increasing size. Conventionally, a computational problem is described as easy or tractable if there exists an efficient algorithm for solving it, one that scales polynomially with input size (for an input of size $n$, as $O(n^k)$ for some $k$). Otherwise, the problem is hard. This distinction is admittedly a rough one: for reasonable problem sizes, an ``inefficient'' algorithm scaling exponentially as $O(1.0001^n)$ would be faster than an ``efficient'' $O(n^{1000})$ algorithm. Nevertheless, this convention has proven useful because, in practice, polynomially scaling algorithms generally outperform exponential ones.

The class of all problems\footnote{Strictly speaking, decision problems, those with a yes-or-no answer. However, other problems can be recast as decision problems; for example, instead of asking ``What is the ground-state energy of molecule $X$?'' we might ask ``Is the ground-state energy of $X$ less than $E_\mathrm{guess}$?''} that are easy for classical computers (classical Turing machines) is called {\sc p}  \cite{Arora2009}. The class of all problems whose answer can be verified in polynomial time is {\sc np}. For example, even though we don't know how to factor numbers efficiently, factoring is in {\sc np} because we can check the proposed answer efficiently by multiplication. Note that {\sc p} $\subseteq$ {\sc np} because any easy problem can be verified easily.   Whether {\sc p} $=$ {\sc np} is a famously open question; however, it is widely believed that they are not equal, that is, that there are problems in {\sc np} that cannot be solved easily \cite{Aaronson2008}. The hardest among them belong to the class {\sc np}-hard: if any {\sc np}-hard problem can be solved efficiently, then so can any problem in {\sc np}.

Likewise, {\sc bqp} are those problems that are easy for a quantum computer \cite{Watrous2009}. The quantum analogue of {\sc np} is called {\sc qma} and contains those problems easy to check on a quantum computer. In analogy with {\sc np}-hard problems, {\sc qma}-hard contains the hardest problems in {\sc qma}. Shor's factoring algorithm \cite{Shor1997} is significant because it provides an example of a problem in {\sc bqp} which is widely thought (although not proven) to be outside of {\sc p}; that is, a problem believed to be hard on classical computers that is easy for a quantum computer.

\begin{figure}
\begin{center}
\includegraphics[width=10cm]{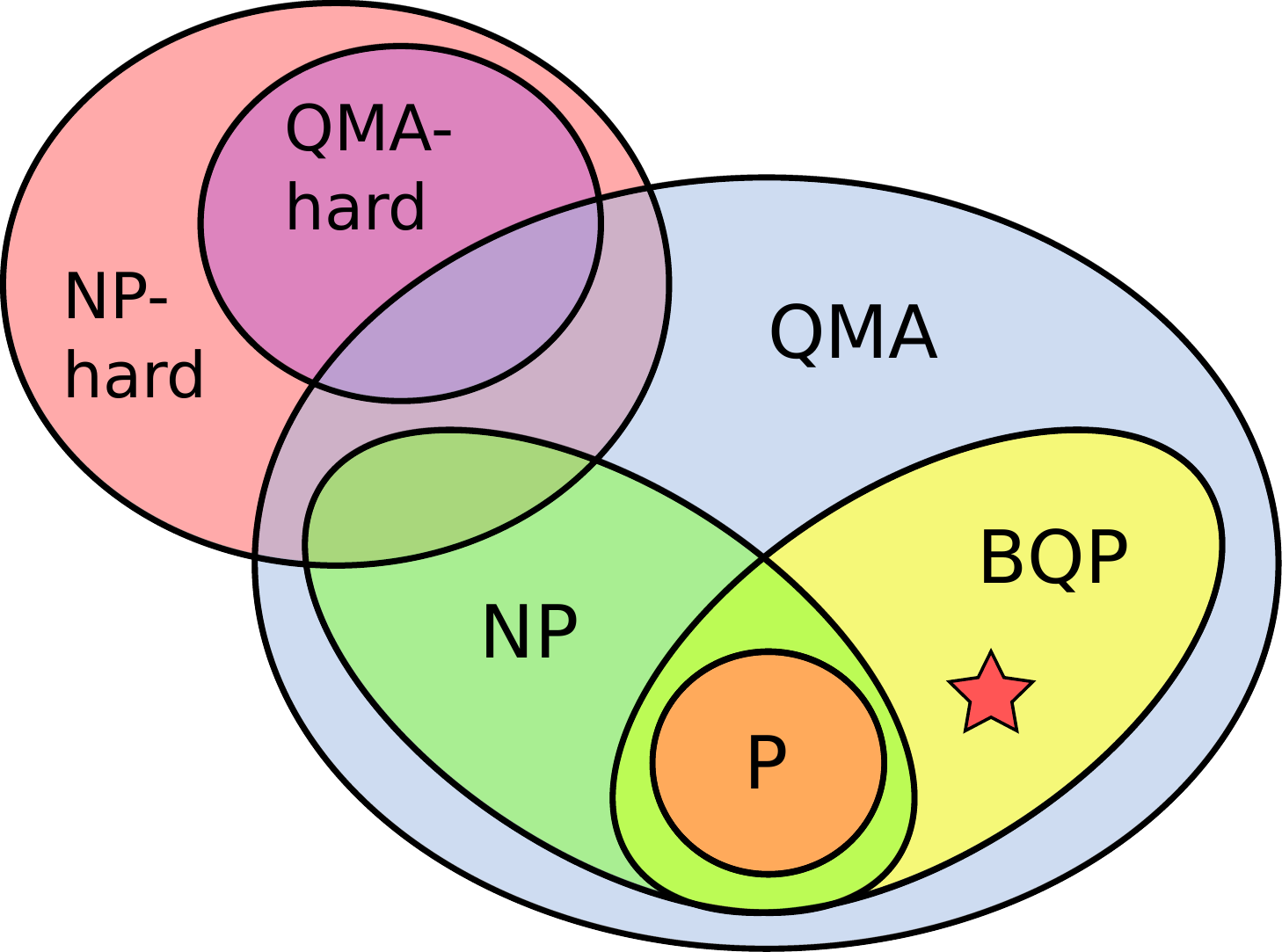}
\end{center}
\caption{\textbf{Computational complexity classes.} Shown are the conjectured relationships between the computational complexity classes discussed in this review. Simulating the time-evolution of chemical systems (denoted by the star) is in {\sc bqp} but widely believed to be outside of {\sc p} (assuming a constant error and simulated time). That is, it is easy on quantum computers, but probably hard---even in principle---on conventional ones.
\label{fig:complexity}}
\end{figure}

The relationships between the complexity classes mentioned above are illustrated in Fig.~\ref{fig:complexity}. In the remainder of this review, we explore the advantages of quantum simulation over its classical counterpart, in part, by situating various simulation tasks in the computational classes illustrated in Fig.~\ref{fig:complexity}.

\section{QUANTUM SIMULATION}\label{sec:qs}

Quantum simulation schemes can be divided into two broad classes. The first is dedicated quantum simulation, where one quantum system is engineered to simulate another quantum system. For example, quantum gases in optical lattices can be used to simulate superfluidity \cite{Greiner2010}. The other, more general, approach is universal quantum simulation, simulating a quantum system using a universal quantum computer\footnote{The terms ``analog'' and ``digital'' have also been used for dedicated and universal quantum simulation, respectively \cite{Buluta2009}.}. Although we will focus on universal quantum simulation because most chemical proposals assume a universal quantum computer, we will mention dedicated simulators where appropriate.

One of the main goals of quantum simulation is to determine the physical properties of a particular quantum system. This problem can usually be conceptualized as involving three steps:
\begin{enumerate}
 \item Initialize the qubits in a state that can be prepared efficiently,
 \item Apply a unitary evolution to this initial state,\footnote{Non-unitary open-system dynamics have been studied as well \cite{Bacon2001}.} and
 \item Read out the desired information from the final state.
\end{enumerate}

We note at the outset that it is not possible to simulate an arbitrary unitary evolution on a quantum computer efficiently. An arbitrary unitary acting on a system of $n$ spins has $2^n\times 2^n$ free parameters, and would require an exponential number of elementary quantum gates to implement. However, in quantum chemistry, it is usually not necessary to simulate arbitrary dynamics, since natural systems aren't arbitrary \cite{Lloyd1996}. Instead, the interactions between, say, molecular orbitals are local---featuring at most $k$-body interactions---and this crucial aspect of their structure can be exploited for their efficient simulation. That is, the Hamiltonian generating the unitary evolution is a sum $H=\sum_i H_i$ of polynomially many terms, each of which acts on at most polynomially many degrees of freedom. A local Hamiltonian generates a time-evolution that can be decomposed into $t/\delta t$ time-steps according to the Lie-Trotter formula,
\begin{equation}\label{eq:trotter}
U=e^{-i\sum_i H_it}\approx \left(\prod_i e^{-iH_i \delta t}\right)^{t/\delta t}.
\end{equation}
The approximation can be improved by increasing the number of time steps or by using higher-order generalizations of this formula \cite{Hatano2005,Berry2007}. Finally, since each factor $e^{-iH_i \delta t}$ acts on only a sub-region of the Hilbert space and can therefore be efficiently simulated, so can a product of polynomially many such factors. Hence, the time it takes to perform the simulation scales polynomially with the simulated time $t$. Most methods of quantum simulation make use of the Trotter decomposition, and we will describe in more detail their applications in chemistry. We will not discuss all the available methods, for which the reader is directed to comprehensive reviews \cite{childs2004,schack_simulationquantum_2006,Brown2010}.

In the following, we describe two ways in which chemical wavefunctions can be encoded on a quantum computer, second- and first-quantization approaches (see Table~\ref{table:comparison} for a comparison). For each approach, we outline the methods of preparing certain classes of initial states and propagating them in time. Afterward, we discuss the methods of measurement of observables and preparation of ground and thermal states, which do not depend essentially on the way the wavefunction is encoded.

{
\renewcommand{\baselinestretch}{1.2}

\begin{table}
\begin{center}
\begin{tabular}{p{1.1in}@{~~~}p{2.5in}p{2.5in}}
\toprule
 & Second-quantized & First-quantized \\

\midrule
Wavefunction encoding &
\begin{minipage}{2.5in}Fock state in a given basis: \vspace{-1mm}\begin{center} \includegraphics[scale=0.4]{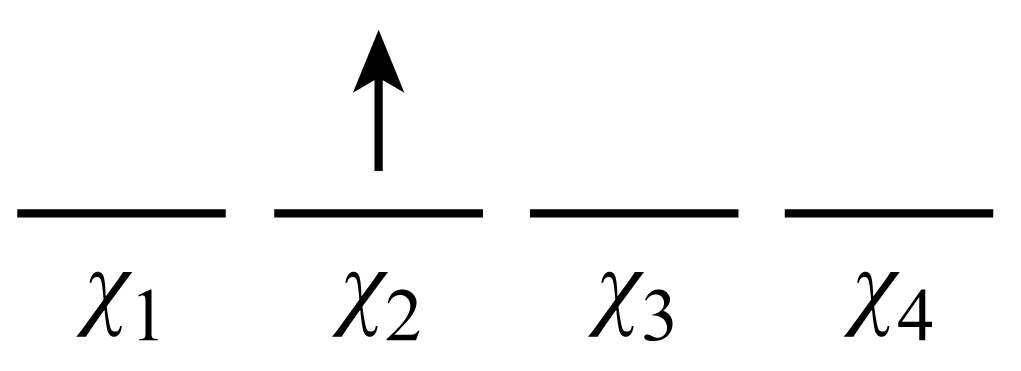}\end{center}\vspace{-3mm}\[\ket{\psi}=\ket{0100}\]\end{minipage} &
\begin{minipage}{2.5in}
\begin{minipage}{1.3in}~\\On a grid of $2^n$ sites per dimension: \[\ket{\psi}=\sum_\mathbf{x} a_\mathbf{x}\ket{\mathbf{x}}\]\end{minipage}\hfill \begin{minipage}{1.1in}\includegraphics[scale=.5]{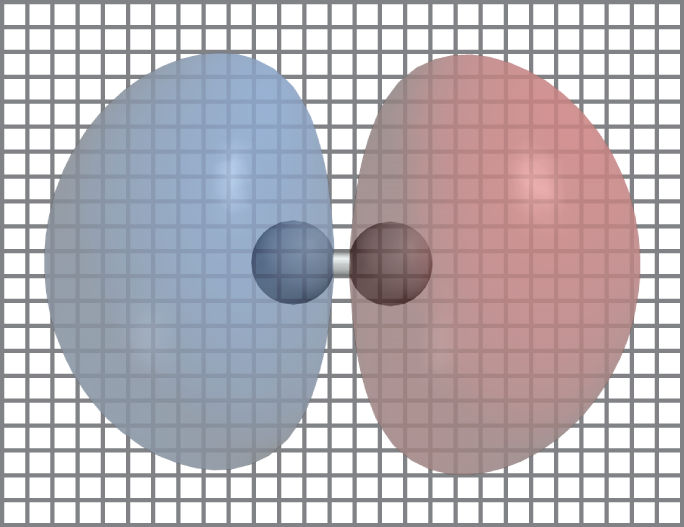} \end{minipage}\end{minipage} \\
\hline

Qubits required to represent the wavefunction &
\begin{minipage}{2.5in}~\\ One per basis state (spin-orbital) \begin{center} \includegraphics[width=2.2in]{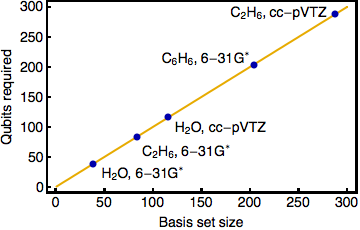} \end{center} \end{minipage} &
\begin{minipage}{2.5in}~\\ $3n$ per particle (nuclei \& electrons) \begin{center} \includegraphics[width=2.2in]{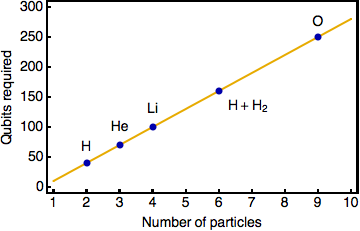} \end{center}\end{minipage} \\
\hline

Molecular Hamiltonian &
\begin{minipage}{2.5in}\[ \sum_{pq}h_{pq}a^\dagger_ p a_q + \frac12 \sum_{pqrs}h_{pqrs}a^\dagger_p a^\dagger_q a_r a_s\] Coefficients pre-computed classically\\ \end{minipage} &
\begin{minipage}{2.5in}\[ \displaystyle \sum_i \frac{p_i^2}{2m_i} + \sum_{i<j} \frac{q_i q_j}{r_{ij}}\] Interaction calculated on the fly \end{minipage} \\
\hline

Quantum gates required for simulation &
\vspace{2mm} $O(M^5)$ with number of basis states &
\vspace{2mm} $O(B^2)$ with number of particles \\
\hline

Advantages &
\begin{minipage}{2.5in}\vspace{4mm}\begin{itemize}
\item Compact wavefunction representation (requires fewer qubits)
\item Takes advantage of classical electronic-structure theory to improve performance
\item Already experimentally implemented \\
\end{itemize}\end{minipage} &
\begin{minipage}{2.5in}\begin{itemize}
\item Better asymptotic scaling (requires fewer gates)
\item Treats dynamics better
\item Can be used for computing reaction rates or state-to-state transition amplitudes
\end{itemize}\end{minipage}\\
\bottomrule
\end{tabular}
\end{center}

\caption{{\bf Comparison of second- and first-quantization approaches to quantum simulation.}
\label{table:comparison}}
\end{table}
}

\subsection{Second quantization}\label{ssec:Second}

We start by considering the purely electronic molecular problem, in which the Born-Oppenheimer approximation has been used to separate the electronic and nuclear motion. The wavefunction of the electrons can be expanded in an orthonormal basis of $M$ molecular spin-orbitals $\{\ket{\chi_i}\}$. Corresponding to this basis are the fermionic creation and annihilation operators $a_i^\dagger$ and $a_i$. There is a very natural mapping between the electronic Fock space and the state of $M$ qubits: Having qubit $i$ in the state $\ket{0}$ (or $\ket{1}$) indicates that spin-orbital $i$ is unoccupied (or occupied).

An important subtlety is that electrons in a molecule, unlike the individually addressable qubits, are indistinguishable. Put differently, while the operators $a_i^\dagger$ and $a_i$ obey the canonical fermionic anticommutation relations,
$\{a_i,a_j^\dagger\}=\delta_{ij}$, the qubit operators that change $\ket{0}$ to $\ket{1}$ and vice versa do not. This problem can be solved by using the Jordan-Wigner transformation to enforce the correct commutation relations on the quantum computer \cite{Ortiz2001, Somma2002, Somma2003, Aspuru2005}. The Jordan-Wigner transformation for this case results in the following mapping between the fermionic operator algebra and the qubit spin algebra:
\begin{subequations}\label{eq:jw}
\begin{align}
a_i^\dagger &\leftrightarrow 1^{\otimes (i-1)} \otimes \sigma^- \otimes \left(\sigma^z\right)^{\otimes (2M-i-1)},\\
a_i &\leftrightarrow 1^{\otimes (i-1)} \otimes \sigma^+ \otimes \left(\sigma^z\right)^{\otimes (2M-i-1)},
\end{align}
\end{subequations}
where $\sigma^-=\ket{1}\bra{0}$ and $\sigma^+=\ket{0}\bra{1}$.

The electronic Hamiltonian in the second-quantized form is
\begin{equation}
H=\sum_{pq}h_{pq}a^\dagger_ p a_q + \frac12 \sum_{pqrs}h_{pqrs}a^\dagger_p a^\dagger_q a_r a_s,
\label{eq:2nd}
\end{equation}
where the spin-orbital indices $p,q,r,s$ each range from 1 to $M$. Here, one-electron integrals $h_{pq}  \equiv \left\langle p \right|\left( {T + V_N } \right)\left| q \right\rangle$ involve the electronic kinetic energy $T$ and the nuclear-electron interaction $V_N$, and the two-electron integrals $h_{pqrs}  \equiv \left\langle {pq} \right|V_e \left| {rs} \right\rangle$ contain the electron-electron interaction term $V_e$. For simulation on a quantum computer, this Hamiltonian is recast into the spin algebra using the Jordan-Wigner transformation, Eq.~\ref{eq:jw}, and the time-evolution it generates is implemented using the Trotter decomposition, Eq.~\ref{eq:trotter}. Note that $H$ contains $O(M^4)$ terms, and each of these terms generates a time evolution of the form
\begin{equation}
U_{pq}=e^{-i h_{pq} a^\dagger_ p a_q \delta t}\quad \mathrm{or} \quad U_{pqrs}=e^{-i h_{pqrs}a^\dagger_p a^\dagger_q a_r a_s \delta t}.
\end{equation}
Each of these operators requires $O(M)$ elementary quantum gates to implement because of the Jordan-Wigner transformation. Since there are altogether $O(M^4)$ terms that need to be implemented separately, the total cost of simulating $H$ scales as $O(M^5)$~\cite{Whitfield2010}.

While any basis $\{\ket{\chi_i}\}$ can be chosen to represent $H$, it is desirable to choose a basis as small as possible that adequately represents the system under study. Electronic structure experience provides for many good starting points, such as the Hartree-Fock basis or the natural orbitals \cite{Davidson1972}. No matter which basis is chosen, a lot of the computation can be carried out on classical computers as pre-processing. In particular, the coefficients $h_{pq}$ and $h_{pqrs}$ can be efficiently pre-computed on classical computers. That way, only the more computationally demanding tasks are left for the quantum computer.

Using the Hartree-Fock basis allows us to use the Hartree-Fock reference state as an input to the quantum computation \cite{Aspuru2005}. A salient feature is that such states are  Fock states, which are easy to prepare on the quantum computer: some qubits are initialized to $\ket{0}$ and others to $\ket{1}$. In fact, any single-determinant state can be easily prepared in this way. Furthermore, it is possible to prepare superpositions of Fock basis states as inputs for the quantum computation. While an arbitrary state might be difficult to prepare, many states of interest, including those with only polynomially many determinant contributions, can be prepared efficiently~\cite{Ortiz2001, Somma2002, Somma2003, Wang2009}. The problem of preparing an initial state that is close to the true molecular ground state is addressed in Sec.~\ref{ssec:Preparing}.

The chief advantage of the second-quantization method is that it is frugal with quantum resources: only one qubit per basis state is required, and the integrals can be pre-computed classically. For this reason, the first chemical quantum computation was carried out in second quantization (see Sec. \ref{sec:expt}). Nevertheless, there are processes, such as chemical reactions, which are difficult to describe in a small, fixed basis set, and for this we turn to discussing first-quantization methods.

\subsection{First quantization}\label{ssec:First}

The first-quantization method, due to Zalka \cite{Zalka1998,Wiesner1996,Kassal2008}, simulates particles governed by the Schr\"odinger equation on a grid in real space\footnote{We also note the method of \cite{Boghosian199830}, which in our terminology is a hybrid between second- and first-quantization methods. It associates a qubit to the occupation of each lattice site.}. For a single particle in one dimension, space is discretized into $2^n$ points, which, when represented using $n$ qubits, range from $\ket{0\ldots 0}$ to $\ket{1\ldots 1}$. The particle's wavefunction can be expanded in this position representation as $\ket{\psi}=\sum_{x=0}^{2^n-1} a_x \ket{x}$. The Hamiltonian to be simulated is
\begin{equation}
H=T+V=\frac{p^2}{2m}+V(x),
\end{equation}
and the resulting unitary can be implemented using the quantum version of the split-operator method \cite{Feit1982,Kosloff1983}:
\begin{equation}
U(\delta t)=e^{-i(T+V)\delta t}\approx U_\mathrm{QFT} e^{-iT\delta t} U^\dagger_\mathrm{QFT} e^{-iV\delta t}.
\end{equation}
The operators $e^{-iV\delta t}$ and $e^{-iT\delta t}$ are diagonal in the position and momentum representations, respectively. A diagonal operator can be easily implemented because it amounts to adding a phase $e^{-i V(x) \delta t}$ to each basis state $\ket{x}$. Furthermore, it is easy on a quantum computer to switch between the position and momentum representations of a wavefunction using the efficient quantum Fourier transform. Therefore, simulating a time evolution for time $t$ involves alternately applying $e^{-iV\delta t}$ and $e^{-iT\delta t}$ with the time steps $\delta t$ chosen to be sufficiently short to secure a desired accuracy. Finally, the scheme can be easily generalized to many particles in three dimensions: a system of $B$ particles requires $3Bn$ qubits, $n$ for each degree of freedom.

The first-quantization method can be applied to many problems. The earliest applications established that as few as 10-15 qubits would be needed for a proof-of-principle demonstration of single-particle dynamics \cite{strini2002error} (later improved to 6-10 \cite{benenti:657}).  The method could also be used to faithfully study the chaotic dynamics of the kicked rotor model \cite{Levi2003}. The first chemical application was the proposal of a method for the calculation of the thermal rate constant \cite{Lidar1999} (see Sec.~\ref{ssec:Measuring}).

We investigated the applicability of the first-quantization method to the simulation of chemical dynamics \cite{Kassal2008}. The simplest approach is to consider all the nuclei and electrons explicitly, in which case the exact non-relativistic molecular Hamiltonian reads
\begin{equation}
H=\sum_i \frac{p_i^2}{2m_i} + \sum_{i<j} \frac{q_i q_j}{r_{ij}},
\end{equation}
where $r_{ij}\equiv|\mathbf{r}_i-\mathbf{r}_j|$ is the distance between particles $i$ and $j$, which carry charges $q_i$ and $q_j$, respectively. As before, the split-operator method can be used to separate the unitaries that are diagonal in the position and momentum bases. Note that a Jordan-Wigner transformation is not required; $H$ preserves permutational symmetry, meaning that if the initial state is properly (anti-)symmetrized (see below), it will stay so throughout the simulation.

Since the Born-Oppenheimer approximation (BOA) has been widely used in quantum chemistry, it might seem extravagant to explicitly simulate all the nuclei and electrons. Nevertheless, the exact simulation is, in fact, faster than using the BOA for reactions with more than about four atoms \cite{Kassal2008}. The reason for this is the need to evaluate the potential $V(\mathbf{r}_1,\ldots,\mathbf{r}_B)$ on the fly on the quantum computer. In the exact case, the potential is simply the pairwise Coulomb interaction; on the other hand, evaluating the complicated, many-body potential energy surfaces that are supplied by the BOA is a much more daunting task, even considering that one can use nuclear time-steps that are about a thousand times longer. That is, exact simulation minimizes arithmetic, which is the bottleneck of the quantum computation; by contrast, the bottleneck on classical computers is the prohibitive scaling of the Hilbert space size, which is alleviated by the BOA.

In order to carry out simulations, it is important to prepare suitable initial states. Zalka's original paper  \cite{Zalka1998} contained a very general state-preparation scheme, later rediscovered~\cite{grover_creating_2002,Kaye2004,kitaev_wavefunction_2008} and improved~\cite{soklakov_efficient_2006}. The scheme builds the state one qubit at a time by performing a rotation (dependent on the previous qubits) that redistributes the wavefunction amplitude as desired.  For example, Gaussian wavepackets or molecular orbitals can be constructed efficiently. We discussed how to combine such single-particle wavefunctions into many-particle Slater determinants, superpositions of determinants, and mixed states in \cite{Ward2009}. In particular, the (anti-)symmetrization algorithm of \cite{Abrams1997}  was improved and used to prepare Slater determinants necessary for chemical simulation. Furthermore, we outlined a procedure for translating states that are prepared in second-quantization language into first-quantized wavefunctions, and vice versa. Techniques for preparing ground and thermal initial states are discussed in Sec.~\ref{ssec:Preparing}.

The first-quantization approach to quantum simulation suffers from the fact that even the simplest simulations might require dozens of qubits and millions of quantum gates \cite{Kassal2008}. Nevertheless, it has advantages that would make it useful if large quantum computers are built. Most importantly, because the Coulomb interaction is pairwise, simulating a system of $B$ particles requires $O(B^2)$ gates, a significant asymptotic improvement over the second-quantized scaling of $O(M^5)$, where $M$ is the size of the basis set.

\subsection{Measuring observables}\label{ssec:Measuring}

We have discussed how to prepare and evolve quantum states on a quantum computer. Information about the resulting state must be extracted in the end; however, full characterization (quantum state tomography) generally requires resources that scale exponentially with the number of qubits. This is because a measurement projects a state into one consistent with the measurement outcome. Because only a limited amount of information can be extracted efficiently, one needs a specialized measurement scheme to extract the desired observables, such as dipole moments, correlation functions, etc.

In principle, an individual measurement can be carried out in any basis. However, since experimental measurement techniques usually address individual qubits, a method is needed to carry out more complicated measurements. In particular, in order to measure an observable $A$, one would like to carry out a measurement in its eigenbasis $\{\ket{e_k}\}$. This is achieved by the phase estimation algorithm (PEA)  \cite{Kitaev1995,Abrams1999}:
\begin{equation}
\frac{1}{\sqrt{N}} \sum_t \ket{t}\ket{\psi} \stackrel{C{-}U}{\longrightarrow}
\frac{1}{\sqrt{N}} \sum_{k,t} {c_k e^{ - iA_k t} } \ket{t}\ket{e_k } \stackrel{QFT}{\longrightarrow}
\sum_k c_k\ket{A_k}\ket{e_k}.
\label{eq:PEA}
\end{equation}
where $c_k  = \braket{e_k }{\psi}$ and $A_k$ are the eigenvalues of $A$; $C{-}U$ is the unitary $U=\exp(-iAt)$ controlled by the ancilla qubits, which are initialized in the state $(1/\sqrt{N}) \sum_t\ket{t}$.  When measuring the ancilla, the eigenvalue $A_k$ will be measured with probability ${\left| {c_k } \right|^2 }$ and, if the eigenstates are non-degenerate, the wavefunction will collapse to the eigenvector $\ket{e_k}$.\footnote{Other methods for eigenvalue measurement include pairing adiabatic quantum evolution with Kitaev's original scheme \cite{Biamonte2010} and applications of the Hellmann-Feynman theorem \cite{Oh2008}.}  For the PEA to be efficient, it must be possible to simulate the pseudo-dynamics $e^{-iAt}$ efficiently. In particular, if we are interested in molecular energies, the observable is the Hamiltonian $H$, and we need to simulate the actual dynamics $e^{-iHt}$ (see Sec.~\ref{sec:qs}). Note that the PEA is closely related to classical algorithms for preparing eigenstates by Fourier analysis of a propagating system \cite{Heller1981,Tannor2006}. As in classical Fourier analysis, the \mbox{(pseudo-)}dynamics must be simulated for longer in order to achieve a higher precision in the $A_k$. More precisely, for a final accuracy of $\epsilon$, the PEA must run for a time $O(1/\epsilon)$ \cite{Nielsen2000,Brown2006}.

Because quantum measurement is inherently random, repeating a measurement on multiple copies of the same system helps in determining expectation values of observables. The central limit theorem implies that measuring $N$ copies of a state results in a precision that scales as $1/\sqrt{N}$ (the standard quantum limit, SQL). For example, repeating the PEA gives an SQL estimate of the coefficients $|c_k|^2$; these can be used to calculate the expectation value $\langle A\rangle=\sum_k |c_k|^2 A_k$, also to the SQL. When entanglement is available, one can achieve precision scaling as $1/N$---this is the Heisenberg or quantum metrology limit \cite{Giovannetti2004}. An algorithm for the expectation values of observables has been proposed that can get arbitrarily close to the Heisenberg limit \cite{Knill2007}.

The first algorithm for measuring a chemical observable was Lidar and Wang's calculation of the thermal rate constant by simulating a reaction in first quantization and using the PEA to obtain the energy spectrum and the eigenstates \cite{Lidar1999}. These values were used to calculate the rate constant on a classical computer by integrating the flux-flux correlation function. We improved on this method with a more direct approach to the rate constant \cite{Kassal2008}. We showed how to efficiently obtain the product branching ratios given different reactant states---if the initial state is a thermal state (see Sec.~\ref{sec:thermal}), this gives the rate constant directly. Furthermore, the method was used to obtain the entire state-to-state scattering matrix. 
A method for reaction rates using a dedicated quantum simulator where artificial molecules are experimentally manipulated, has also been proposed \cite{Smirnov2007}.

More generally, correlation functions provide information about a system's transport and spectroscopic properties. On a quantum computer, the correlation function of any two observables can be estimated efficiently if their pseudo-dynamics can each be simulated efficiently  \cite{Ortiz2001, Somma2003}. The method does not suffer from the dynamic sign problem that plagues classical Monte Carlo methods for sampling correlation functions.
An alternative approach is the measurement of correlation functions using techniques of linear-response theory \cite{Terhal2000}.

Molecular properties such as the dipole moment or the static polarizability are also of chemical interest. They are derivatives of the molecular energy with respect to an external parameter, such as the electric field. We showed how to calculate them \cite{Kassal2009} using the PEA and the quantum gradient algorithm \cite{Jordan2005}. The algorithm is insensitive to the dimensionality of the derivatives, an obstacle to classical computers. For example, the molecular gradient and Hessian can be computed---and used to optimize the geometry---with a number of energy evaluations independent of system size.

\subsection{Preparing ground states and thermal states}\label{ssec:Preparing}

In Secs.~\ref{ssec:Second} and~\ref{ssec:First}, we discussed the preparation of various initial states for quantum simulation. We postponed discussing the preparation of ground and thermal states because of subtleties to which we now turn.

\subsubsection{Ground state preparation by phase estimation}\label{ssec:ground}

A large part of quantum chemistry is concerned with the calculation of ground-state properties of molecules, making it desirable to prepare such states on a quantum computer. In the previous section, we described how the PEA can be used to measure a quantum state in the eigenbasis of a Hermitian operator. This suggests a method for preparing a ground state: measuring in the eigenbasis of the Hamiltonian will project a state $\ket{\psi}$ to the ground state $\ket{g}$ with probability $\left| \left\langle \psi | g \right\rangle \right|^2$.

The problem, therefore, is to prepare a state close to the ground state, from which we can project the ground-state component. Choosing a random state $\ket{\psi_\mathrm{rand}}$ is bound to fail, since the overlap  is expected to be exponentially small in the number of qubits $n$: $\left\langle \psi_\mathrm{rand} | g \right\rangle \sim 2^{-n}$. This means that one would have to repeat the PEA exponentially many times before chancing upon the ground state.

Methods of quantum chemistry can be used to improve the overlap.  We studied the ground-state preparation of H$_2$O and LiH in second quantization, based on the Hartree-Fock (HF) approximation \cite{Aspuru2005}. The goal was to prepare the ground state of the full configuration interaction (FCI) Hamiltonian, so that its energy could be read out by the PEA, thus solving the electronic structure problem. Since these molecules were considered at equilibrium geometries, the HF guess was sufficient for the algorithm to estimate the ground-state energies of these molecules with high probability. The overlap can be further improved by choosing a more sophisticated approximation method such as a multi-configuration self-consistent field (MCSCF) wavefunction \cite{Wang2008}.

Alternatively, the overlap can be increased using adiabatic quantum computation (Sec.~\ref{sec:adiabatic_intro}). We applied adiabatic state preparation (ASP) to the case of the hydrogen molecule H$_2$ in the STO-3G basis at various bond lengths \cite{Aspuru2005}. As the bond length increases, the HF state has decreasing overlap with the exact state, reaching 0.5 at large separations. ASP works by preparing the ground state of the HF Hamiltonian and then slowly changing to the FCI Hamiltonian. The speed of the variation of the Hamiltonian is limited by the energy gap between the ground state and the first excited state. In the case of H$_2$, this method allowed the preparation of the FCI ground state with a high fidelity.

Procedures similar to ASP have been proposed to study low-energy states of some toy models in physics \cite{Oh2008} and superconductivity \cite{Wu2002}. It is also possible to encode a thermal state into the ground state of a Hamiltonian \cite{Aharonov2008,Somma2007}, offering a way to prepare a thermal state, a problem further discussed in the next section.

\subsubsection{Thermal state preparation}\label{sec:thermal}

While not often a subject of quantum-chemical calculations, the thermal states are of significance because they can be used to solve many problems, ranging from statistical mechanics to the calculation of thermal rate constants. Classical algorithms typically rely on Markov chain Monte Carlo (MCMC) methods, which sample from the Gibbs density matrix, $\rho  = e^{ - \beta H} /Z$, where $Z$ is the partition function. The challenge is that it is generally impossible to sample from the eigenstates $\ket{e_k}$ of a certain Hamiltonian $H$ if they are not pre-determined (which is often more challenging).

With a quantum computer, assuming the PEA can be efficiently implemented, we can prepare the thermal state of any classical or quantum Hamiltonian from a Markov chain constructed by repeating a completely positive map \cite{Terhal2000}. A limitation of that approach is that the Metropolis step can make too many transitions between states of very different energies, sometimes leading to a slow convergence rate of the resulting Markov chain. This issue was addressed by building up the Markov chain by applying random local unitary operations \cite{Temme2009}. The resulting operation is a Metropolis-type sampling for quantum states; although the underlying Markov chain is classical in nature, performing it on a quantum computer provides the benefit of being able to use the PEA without explicitly solving the eigenvalue equations. However, quantum computers can implement Markov chains corresponding to thermal states of classical Hamiltonians with a quadratic speed-up \cite{Szegedy04, Somma2008, Wocjan08, Poulin2009,Poulin2009a,chiang2010quantum}.

Zalka's state preparation algorithm (see Sec.~\ref{ssec:First}) is applicable to preparing the coherent encoding of thermal states (CETS) $\left| {\psi _\mathrm{th} } \right\rangle $,
\begin{equation}
\left| {\psi _\mathrm{th} } \right\rangle  = \sum\limits_k {\sqrt {e^{ - \beta E_k } /Z} } \ket{e_k}\ket{e_k} ,
\end{equation}
which is equivalent to the Gibbs density matrix, $\rho _\mathrm{th}  = \sum\nolimits_k {e^{ - \beta E_k } /Z} \ket{e_k}\bra{e_k}$, if one register is traced out. If the eigenstates and eigenvalues are known, it is possible to construct the CETS directly \cite{Ward2009}. On the other hand, combining ideas from belief-propagation \cite{Mezard2009} and quantum amplitude amplification \cite{Kaye2007}, we were able to construct the CETS of classical Hamiltonians with a quadratic quantum speedup \cite{Yung2010}.

Lastly, a thermal state can be prepared by modeling the physical interaction with a heat bath \cite{Zalka1998,Terhal2000}. However, the computational cost of these methods is not well understood.

\subsubsection{QMA-hardness and future prospects}

Unfortunately, the procedures for ground- and thermal-state preparation outlined above are not fully scalable to larger systems. A simple way to see this is to imagine a system composed of $N$ identical, non-interacting molecules. Even if one molecule can be prepared with a ground-state overlap of $1-\epsilon$ by any method, the fidelity of the $N$-molecule state will be exponentially small,  $(1-\epsilon)^N $ \cite{Kohn1999}. ASP would fail when the energy gap got so small that the Hamiltonian would have to be varied exponentially slowly.

More generally, there are broad classes of Hamiltonians for which finding the ground state energy (and therefore also a thermal state) is known to be {\sc qma}-hard, that is, most likely hard even on a quantum computer (see Sec.~\ref{ssec:complexity}) \cite{kitaev_classical_2002,aharonov_quantum_2002,kempe2004,Oliveira2005,schuch_computational_2009}. Nevertheless, the scaling of the ground- and thermal-state energy problems for chemical systems on a quantum computer is an open question. It is possible that algorithms can be found that are not efficient for all {\sc qma}-hard Hamiltonians, but nevertheless succeed for chemical problems. 

\section{OPTIMIZATION WITH ADIABATIC QUANTUM SIMULATION}\label{sec:Adiabatic}

\begin{figure}
	\begin{center}
	\includegraphics[width=10cm]{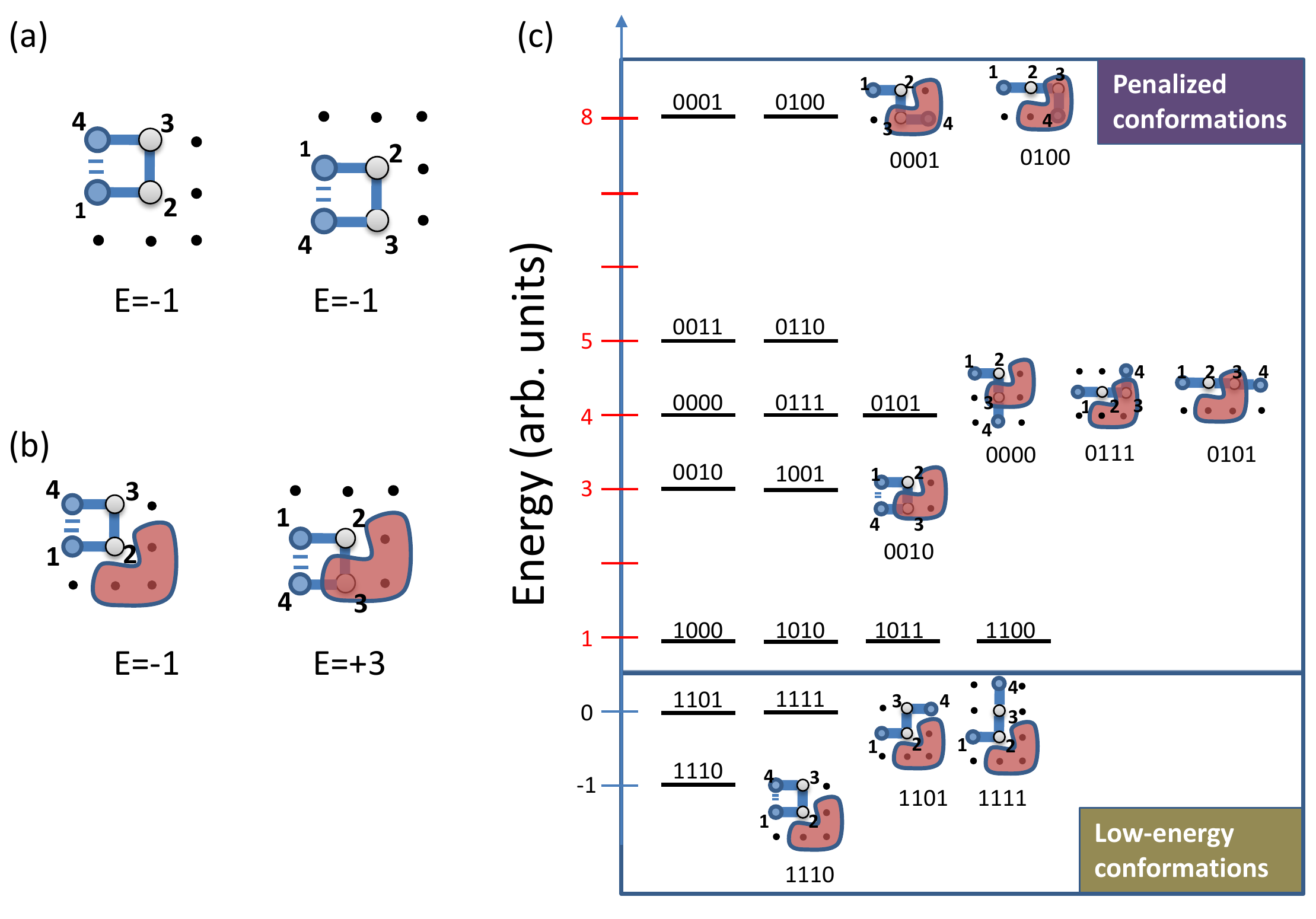}
	\end{center}
	\caption{\label{fig:AQC}\textbf{Lattice folding using quantum adiabatic hardware:} In vivo folding of proteins involves the assistance of molecular chaperone proteins whose main function is to assist the folding of the newly synthesized polypeptide. (a) In \textit{vacuo}, the four-amino-acid peptide could fold clockwise or counterclockwise, depending if the third amino acid moves downwards or upwards, respectively. (b) The chaperone molecule (represented as the pink region) obstructs the third amino acid from moving downward, making the counterclockwise folding the only global minima in the energy landscape. (c) \textit{Energy landscape} (Eq.~\ref{eq:E4qubit4local}): Each overlap of the amino acids with the chaperone raises the free energy by four units, while overlap among amino acids in the chain raises the free energy by two units. The four binary variables encode the direction of the third and fourth bond of the peptide (see text). The quantum adiabatic hardware found the right solution 78 \% of the times~\cite{Perdomo-OrtizAQCExp2010}.}
\end{figure}

We describe the use of quantum computers to solve classical optimization problems related to chemistry and biology. This class of problems plays an important role in fields such as drug design, molecular recognition, geometry optimization, protein folding, etc. ~\cite{hartmann2004,floudas2000}.

Of all the models of quantum computation, adiabatic quantum computation (AQC) is perhaps the best suited for dealing with discrete optimization problems. As explained in Sec.~\ref{sec:adiabatic_intro}, the essential idea behind AQC is to encode the solution to a computational problem in a (final) Hamiltonian ground state which is prepared adiabatically.

Although final Hamiltonians have been proposed for various problems related to computer science~\cite{farhi2000,hogg03,neven2008,neven2009,choi2010}, only recently we derived constructions~\cite{Perdomo2008} for problems of chemical interest such as the lattice heteropolymer problem~\cite{pande_heteropolymer_2000,mirny2001,kolinski1996}, an {\sc np}-hard problem~\cite{hart_robust_1997}.  It can be used as a model of protein folding \cite{dill_protein_2008}, one of the cornerstones of biophysics. Note that the quantum-computational implementation of the protein folding problem does not assume that the protein is treated quantum mechanically. Instead, the quantum computer is being used as a tool to solve the classical optimization problem. In the lattice folding problem, the sequence of amino acids is coarse-grained to a sequence of beads (amino acids) connected by strings (peptide bond). This chain of beads occupies points on a two- or three-dimensional lattice; a valid configuration (fold) is a self-avoiding walk on the lattice and its energy is determined by the interaction energies among amino acids that are non-bonded nearest neighbors in the lattice. The hydrophobic-polar (HP) model~\cite{LAU1989} is the simplest realization of this problem. The amino acids are broken into two groups, hydrophobic (H) and polar (P). Whenever two non-bonded H amino acids are nearest neighbors in the lattice, the free-energy of the protein is reduced by one unit of energy, $E_{HH}=-1$. The remaining interactions do not contribute to the free energy $E_{HP}=E_{PP}=0$. The lattice folding problem consists in finding one of more folds that minimize the free energy of the protein. By the thermodynamic hypothesis~\cite{Epstein1963}, such fold(s) correspond to the conformation of the native conformation(s) of the protein.

The theory behind the quantum-computational implementation of lattice folding is guided by the proposed quantum adiabatic platform on superconducting qubits~\cite{kaminsky2004}. This scheme is designed to find solutions to the problem,
\begin{equation}\label{eq:QUBO}
E(\mathbf{s})  = - \sum_{i} h_{i} s_i  + \sum_{j>i} J_{ij} s_{i} s_{j},
\end{equation}
where $\abs{h_i} \le 1$, $\abs{J_{ij}} \le 1$, and $s_i = \pm 1$. Given a set of $\{h_i\}$ and the interaction matrix $\{J_{ij}\}$, the goal is to find the assignment $\mathbf{s^*} = s^*_1 s^*_2 \cdots s^*_N$, that minimizes $E(\mathbf{s})$.

The time-dependent Hamiltonian is chosen to be,
\begin{equation}\label{h-AQO}
H(\tau)  = A(\tau) H_i + B(\tau) H_f, \quad \quad \tau= t/t_{run},
\end{equation}
where $H_i$ has a simple-to-prepare ground state and $H_f = - \sum_{i} h_{i}\sigma_{z}^{i}  + \sum_{j>i} J_{ij}\sigma_{z}^{i} \sigma_{z}^{j}$, with $\sigma_{z}^{i}$ denotes the Pauli matrix acting on the $i$th qubit, and $t_{run}$ is the running time. The time-dependent functions $A(\tau)$ and $B(\tau)$ are such that $A(0) \gg B(0) $ and $A(1) \ll B(1)$.   Therefore, at the beginning (end) of the simulation, the ground state corresponds to the ground state of $H_i$ ($H_f$). Note that, as desired, $\ket{\mathbf{s^*}} \equiv \ket{s^*_1, s^*_2, \cdots, s^*_N}$ is the ground state of $H_f$. Measurement of this final state provides the solution to our problem.

The theoretical challenge is to map the lattice folding free energy function into the form of Eq.~\ref{eq:QUBO}~\cite{Perdomo2008,Perdomo-OrtizTURNmapping2010}. In two dimensions, we use two binary variables determining the direction of each bond between two amino acids (beads). If a particular bond points upwards, we write ``11"; if it points downwards, leftwards or rightwards, we write ``00", ``10", or ``01", respectively. For an $N$-amino-acid protein, we need two binary variables for each of the $N-1$ bonds. Fixing the direction of the first bond reduces the number of variables to $\ell =2(N-2)$ binary variables. Any possible $N$-bead fold can be represented by the string of binary variables of the form $0 1 q_1 q_2 \cdots q_{\ell-1} q_{\ell}$, where we set the direction of the first bond to be right (``01").

As an example, the free energy function~\cite{Perdomo-OrtizTURNmapping2010} associated with the folding of a four-amino-acid peptide assisted by a ``chaperone" protein (see Fig.~\ref{fig:AQC}) is
\begin{equation}\label{eq:E4qubit4local}
\begin{split}
E(\boldsymbol{q})  &= 4 - 3 q_{1} + 4 q_{2} - 4 q_{1} q_{2} - q_{3} + q_{1} q_{3} - 2 q_{2} q_{3} +
 4 q_{4} - 2 q_{1} q_{4} \\&  - 8 q_{2} q_{4} + 5 q_{1} q_{2} q_{4} -
 2 q_{3} q_{4} + 5 q_{2} q_{3} q_{4} - q_{1} q_{2} q_{3} q_{4}.
\end{split}
\end{equation}
By substituting values for the four binary variables defining the directions of the second ($q_1 q_2$) and third ($q_3 q_4$) bonds, we can verify that the 16 assignments provide the desired energy spectrum (Fig.~\ref{fig:AQC}). Eq.~\ref{eq:E4qubit4local} is not in the form of Eq.~\ref{eq:QUBO}. We converted this energy function from its quartic form to a quadratic form, using two extra ancilla binary variables~\cite{Perdomo-OrtizTURNmapping2010}. After the substitution $q_i \equiv \frac{1}{2} (1 - s_i)$, the free energy function now resembles that of Eq.~\ref{eq:QUBO}. An early experimental realization is described in Sec.~\ref{ssec:squids}.

Note that solving the HP model is {\sc np}-hard~\cite{Berger1998,Crescenzi1998,hart_robust_1997}. AQC is equivalent to the circuit model, so it is unlikely able to solve {\sc np}-hard problems in polynomial time (see Sec.~\ref{ssec:complexity}). Real world problems (and the instances defining biologically relevant proteins) are not necessarily structureless. Taking advantage of the structure or information about a particular problem instance is one of the ideas behind new algorithmic strategies~\cite{amin2009,farhi2009,PerdomoSAQC2010}. An example is to introduce heuristic strategies for AQC by initializing the calculation with a educated guess~\cite{PerdomoSAQC2010}.

\section{EXPERIMENTAL PROGRESS}\label{sec:expt}

Experimental quantum simulation has rapidly progressed \cite{Buluta2009,Brown2010} since the early simulation of quantum oscillators using nuclear magnetic resonance (NMR)~\cite{Somaroo1999}. Here we review chemical applications of available quantum-computational devices.

\subsubsection*{Quantum optics}

\begin{figure}
	\begin{center}
	\includegraphics[width=10cm]{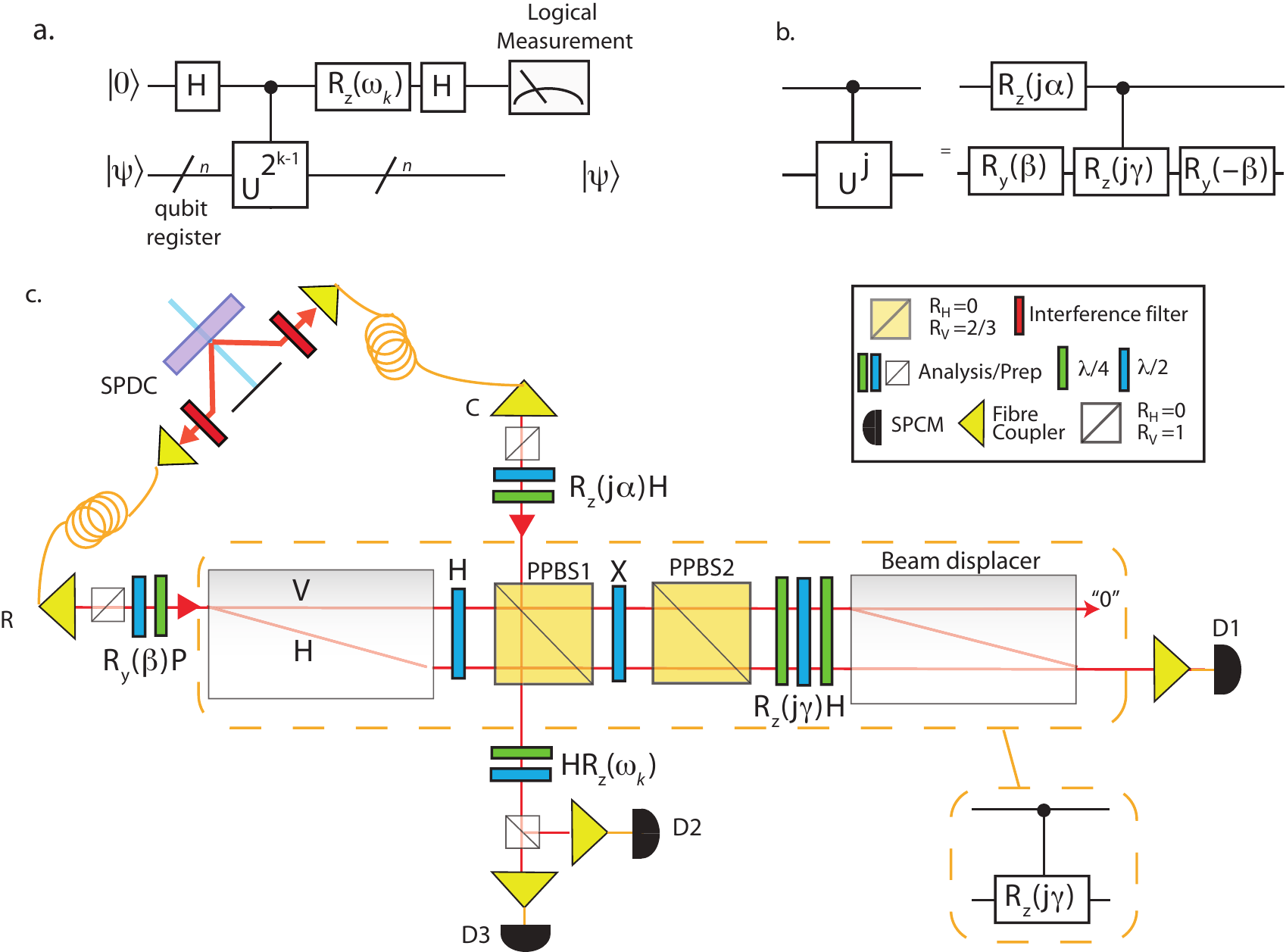}
	\includegraphics[width=9cm]{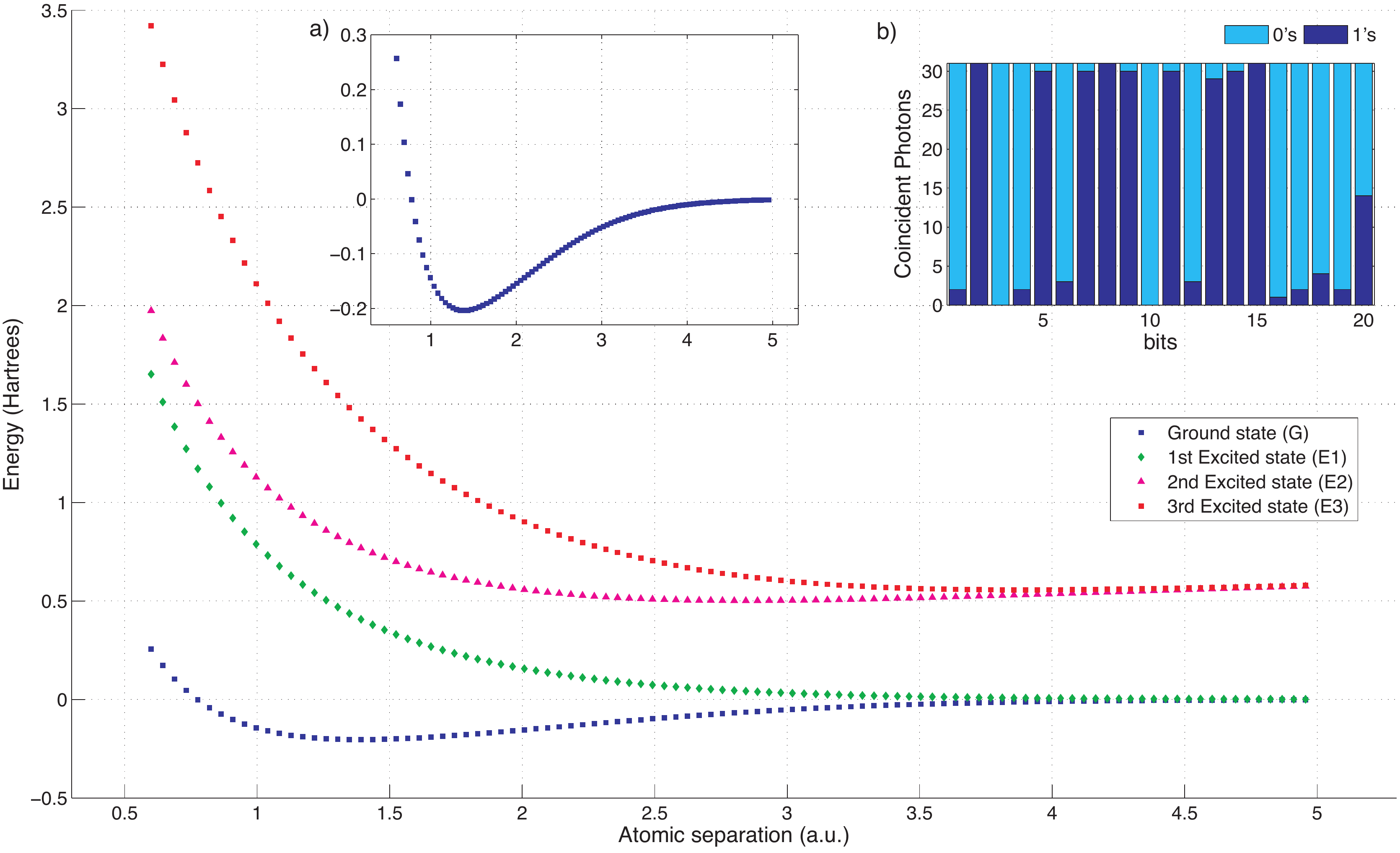}
	\end{center}
	\caption{\textbf{Experimental simulation of the H$_2$ molecule on a linear-optical quantum computer \cite{Lanyon2010}.} a) Two-qubit iterative version of the phase estimation algorithm for evaluating molecular energies. b) Decomposition of the algorithm into gates. c) The layout of the optical elements used to implement the quantum gates on photonic polarization qubits. d) The computed potential energy surfaces of the H$_2$ molecule in a minimal basis set. The results are the exact (in the basis) full configuration interaction energies, to 20 bits of precision.
	\label{fig:experiment}}
\end{figure}

On an optical quantum computer, various degrees of freedom of single photons, such as polarization or path, are used to encode quantum information \cite{OBrien2007,Kok2007}. This architecture was used for the first quantum simulation of a molecular system, a minimal basis model of the hydrogen molecule H$_2$ \cite{Lanyon2010}.  Qubits were encoded in photon polarization, while two-qubit gates were implemented probabilistically using linear-optical elements and projective measurement. The minimal-basis description of H$_2$ used two spin-orbitals per atom. Since the FCI Hamiltonian is block-diagonal with $2\times 2$ blocks, two qubits sufficed for the experiment: one for storing the system wavefunction, and one for the readout of the PEA. The PEA was implemented iteratively, extracting one bit of the value of the energy at a time. Twenty bits of the energy were obtained, and the answer was exact within the basis set. Fig.~\ref{fig:experiment} describes the experiment and the potential energy surfaces that were obtained.

\subsubsection*{Nuclear magnetic resonance}

Nuclear spins can serve as qubits, being addressed and read-out using an NMR spectrometer \cite{baugh_quantum_2007}.  The first experimental quantum simulation, of a harmonic oscillator, was performed using NMR~\cite{Somaroo1999}. The platform has since been used to simulate a number of model systems~\cite{negrevergne_liquid-state_2005,Brown2006,Yang2006,Peng2005}, leading up to the recent simulation of H$_2$ \cite{Du2010}. The H$_2$ experiment used $^{13}$C-labeled chloroform, in which the carbon and hydrogen nuclear spins form two qubits.  The experiment achieved 45 bits of precision (15 iterations of PEA, 3 bits per iteration) in the ground state energy. Adiabatic state preparation (Sec.~\ref{ssec:Preparing}) was implemented for various bond distances.

\subsubsection*{Superconducting systems}\label{ssec:squids}

The circulating current (clockwise or counterclockwise) flowing in a micron-sized loop of a superconductor can be used as a qubit~\cite{You2005,levi2009}. Examples of applications based on superconducting qubits include the tailor-made generation of harmonic oscillator states~\cite{hofheinz2009} and the implementation of the Deutsch-Jozsa and Grover quantum algorithms~\cite{dicarlo2009}. Recently, the free energy function discussed in Sec.~\ref{sec:Adiabatic} for the four-amino-acid peptide assisted by a chaperone protein (see Fig.~\ref{fig:AQC}) has been experimentally realized~\cite{Perdomo-OrtizAQCExp2010}. A microprocessor consisting of an array of coupled superconductor qubits has been used to implement the time-dependent Hamiltonian in Eq.~\ref{h-AQO}, with $H_i  \propto \sum_i \sigma_{x}^{i}$ as the initial Hamiltonian~\cite{kaminsky2004,johnson2010,harrisPRB2010}. The quantum hardware operating at a temperature of 20 mK found the correct solution with a probability of 78\%. Characterization of this device is currently underway \cite{harris2009,berkley2009,harris2010,lanting2010}.

\subsubsection*{Trapped ions}

Qubits can also be encoded in the electronic states of cold trapped ions, offering one of the most controllable systems available today \cite{cirac_quantum_1995,Johanning2009,blatt_entangled_2008}. This platform has already produced sophisticated simulations of physical systems \cite{friedenauer_simulatingquantum_2008,Kim2010,Gerritsma2010}, but chemical applications are still to come.

\section{CONCLUSIONS}\label{sec:concl}

We have outlined how a quantum computer could be employed for the
simulation of chemical systems and their properties, including correlation
functions and reaction rates. A method
for lattice protein folding was also discussed.
Although we focused on the adiabatic and circuit models, these are not
the only universal models of quantum computation and it may be
possible to make further algorithmic progress with models such as
topological quantum computing
\cite{kitaev_fault-tolerant_2003,nayak_non-abelian_2008}, one-way
quantum computing
\cite{raussendorf_one-way_2001,raussendorf_measurement-based_2003},
and quantum walks \cite{kempe_quantum_2003,Childs2009,Lovett2010}.

We reported on the first experiments relevant to chemistry and we
expect more to come in the near future.  With recent technological
advances, there are many prospects for the future of quantum
simulation.  However, as more qubits are added to experiments, more
effort will be needed to control decoherence, since error correction
procedures \cite{Nielsen2000} might not be sufficient in practice due
to the spatial and temporal overheads required
\cite{Clark2009}. Instead, it may be possible to build resilient
quantum simulators or to incorporate the noise into the simulation.

Although practical quantum computers are not available yet, quantum
information theory has already influenced the development of new
methods for quantum chemistry. For instance, density matrix
renormalization group has been extended using quantum information, and
its applications to chemistry have been vigorously pursued
\cite{Chan2008}. By studying the simulation of chemical systems on
quantum computers, we can also expect new insights into the complexity
of computing their properties classically.

In analogy to classical electronics, one could say that, as of 2010, the
implementation of quantum information processors is in the vacuum-tube
era. A development parallel to that  of the transistor would allow for
rapid progress in the capabilities of quantum information processors.
These larger devices would allow for routine execution of exact,
non-adiabatic dynamics simulations, as well as full-configuration
interaction calculations of molecular systems that are intractable
 with current classical computing technology.

\subsection*{Summary Points}
\begin{itemize}
	\item A universal quantum computer can simulate chemical systems more efficiently (in some cases exponentially so) than a classical computer.
	\item Preparing the ground state of an arbitrary Hamiltonian is a hard ({\sc qma}-complete) problem. However, the ground state of certain chemical Hamiltonians can be found efficiently using quantum algorithms.
	\item Simulation of quantum dynamics of physical systems is in general efficient with a quantum computer.
	\item Properties of quantum states can be obtained by various measurement methods.
	\item Classical optimization problems, such as lattice protein folding, can be studied by means of the adiabatic quantum-computational model.
	\item Quantum simulation for chemistry has been experimentally realized in quantum optics, nuclear magnetic resonance, and superconducting devices.
\end{itemize}

\subsection*{Future Issues}
\begin{itemize}
\item Developing quantum simulation methods based on alternative models of quantum computation is an open research direction.
\item Dedicated quantum simulators built so far are mostly for
  simulating condensed matter systems. It is desirable to make
  experimental progress on simulating chemical systems.
\item Decoherence is currently the major obstacle for scaling up the
  current experimental setups. Progress in theoretical and
  experimental work is needed to overcome decoherence.
\item We have not covered methods of quantum error correction, which
  will be important for large scale simulations.
\end{itemize}


\bibliographystyle{abbrv}
\bibliography{QuanSim}

\end{document}